\documentclass[final,5p,times,twocolumn,authoryear]{elsarticle}
\usepackage{amsmath,amssymb,amsthm,mathtools,graphicx,booktabs,microtype,placeins}
\journal{Automatica}
\newtheorem{assumption}{Assumption}

\newtheorem{theorem}{Theorem}
\theoremstyle{remark}
\newtheorem{remark}{Remark}
\theoremstyle{plain}

\newcommand{\R}{\mathbb{R}}
\begin{document}
\begin{frontmatter}
\title{Delay-Enabled Adaptive Learning of Time-Varying Parameters}
\author[uri]{Chengzhi Yuan}
\ead{cyuan@uri.edu}
\author[gso]{Mingxi Zhou}
\ead{mzhou@uri.edu}
\address[uri]{Department of Mechanical, Industrial and Systems Engineering, University of Rhode Island, Kingston, RI 02881, USA}
\address[gso]{Graduate School of Oceanography, University of Rhode Island, Narragansett, RI 02882, USA}
\begin{abstract}
This paper proposes a novel adaptive learning method  for unknown time-varying parameters by deliberately incorporating a delay into the parameter update. Specifically, a conventional adaptive estimator is augmented with only one additional dynamic state, which estimates the parameter variation rate from the difference between current and previous parameter estimates. The method fundamentally changes the influence of time variation on estimation performance: the resulting estimation error is governed by the second derivative of the unknown parameter rather than its first derivative. Under persistent excitation, a small-gain analysis proves exponential stability of the complete estimation-error system and establishes explicit bounds for the parameter, state-prediction, and parameter-rate errors. In particular, accurate learning of linearly time-varying parameters and their unknown rates can be guaranteed with errors converging exponentially fast to zero.  Numerical comparisons with conventional adaptive estimation and a recent time-varying parameter estimator demonstrate superior learning performance in the considered examples. The proposed delay-enabled adaptive estimator requires neither state derivatives nor any assumed model of parameter variation, and its implementation consists of one delayed difference and a first-order filter. These results thus provide a direct extension of adaptive learning from constant to time-varying parameters without sacrificing the simplicity of the conventional adaptive estimation mechanism.
\end{abstract}
\end{frontmatter}

\section{Introduction}\label{sec:intro}

Adaptive control has a long and mature history. Much of its classical theory was built around systems containing unknown but constant parameters, for which the central questions are whether the closed loop remains stable, whether the parameter estimates remain bounded, and, under persistent excitation (PE), whether the unknown constants can be identified accurately \citep{ioannou}. Over several decades, this framework has led to a remarkably complete body of theory for linear and nonlinear systems, robust adaptation, parameter convergence, and learning from measured data. In this classical setting, however, the object to be learned is fundamentally static: the parameter may be unknown, but its value does not vary with time.

Extending this theory to parameters that vary with time has proved substantially harder, and the literature remains far smaller and more fragmented than its constant-parameter counterpart. Early work established stability and adaptation results for slowly or continuously time-varying plants \citep{middleton}; robust and least-squares identification methods were later developed for slowly varying parameters and bounded disturbances \citep{mathelin,joensen}. Polynomial-in-time approximations and interval-wise resetting were subsequently used to estimate more general time-varying parameters \citep{na2015}. More recent approaches include directional forgetting \citep{leeDF}, filtered parameter-error information for nonlinear systems \citep{na2021}, $L_2$-gain tuning that treats parameter variation as a disturbance \citep{ruedaL2}, and local polynomial approximation combined with dynamic regressor extension and mixing (DREM) \citep{aranovskiy,diget}. Time-varying adaptation gains have also been used to obtain parameter-error bounds under finite or persistent excitation \citep{gaudio2022}, while associative-memory-based recursive estimation has been investigated for varying parameters modeled as unknown functions of measurable variables \citep{alegria2022}.  These advances demonstrate that time-varying parameter estimation is possible, but they also reveal how difficult the problem becomes once the unknown quantity is no longer static. Existing approaches
typically rely on additional filtering or reconstruction, forgetting mechanisms, disturbance-attenuation design, or an explicitly prescribed local model of parameter variation with selected basis
functions and approximation windows. Such methods extend the reach of classical adaptation, but the resulting architectures are often considerably more involved, and the variation of unknown parameters is handled through auxiliary modeling or tracking machinery rather than through a simple learning mechanism intrinsic to the adaptive law itself.

This paper takes a different route. The idea is motivated by the ``using delay for control'' perspective of \citet{fridmanDelay}, where a deliberately introduced delay is treated not merely as an adverse phenomenon but as a design resource. In this paper, instead of introducing delay into the plant or controller, a known delay is deliberately embedded inside the adaptive learner. The difference between the present parameter estimate and its value a short time earlier contains information about the parameter change over that interval. A single first-order dynamic state uses this difference to estimate the instantaneous parameter variation rate. Meanwhile, the standard adaptive predictor continues to use current measured data. This leads to a particularly parsimonious adaptive learning architecture for unknown time-varying parameters. No derivative of the unknown parameter is measured, no differential model for its variation is imposed, and no local polynomial basis, resetting rule, or moving-window model is required. The learner uses only one delay, one finite difference, and one additional rate-estimation state. Yet this small architectural change gives the adaptive mechanism a qualitatively new capability: it does not merely estimate the current value of an unknown time-varying parameter, but also learns how that parameter is changing.

In addition, conventional estimators must continually respond to the unknown first derivative of the parameters. In the proposed estimator, that first-order variation is learned by the additional state, and the remaining external terms depend on the second derivative. Therefore, the proposed method can learn parameters that vary linearly with time, including their constant but unknown rates, with vanishing estimation error. For more general smoothly varying parameters, the error is bounded directly in terms of how rapidly the parameter rate itself changes. This improvement is obtained without measuring parameter derivatives or introducing a model for the unknown parameter motion.

The proposed scheme is supported by a unified stability analysis. Specifically, the resulting complete estimation-error system is formed as an interconnection between a standard linear time-varying subsystem and a first-order delay subsystem. Exponential stability of the former is established under the PE condition, while the interaction between the two subsystems is analyzed using the small-gain theorem \citep{vanderschaft2017}. This gives a concise stability condition and an explicit error bound for given learner gains, filter time constant, and delay.

The main contributions of this paper are threefold. First, we introduce a novel delay-enabled adaptive learning law that estimates both unknown time-varying parameters and their rates with one additional rate-estimation state. Second, we prove exponential stability and quantitative error bounds that depend on the second, rather than first, derivative of the unknown parameters; linearly time-varying parameters are learned exactly as a direct consequence. Third, we establish a small-gain stability test for the complete error system and evaluate the method against representative estimators on a nonlinear system. Together, these results substantially broaden the class of unknown parameter variations that can be learned using a simple adaptive structure.

The remainder of the paper is organized as follows. Section~\ref{sec:problem} states the problem and objective. Section~\ref{sec:results} presents the proposed learning law and stability analysis. Sections~\ref{sec:sim} and \ref{sec:conclusion} report the numerical studies and conclusions, respectively.

\emph{Notation.} $\R$ is the set of real numbers, $I_p$ is the $p\times p$ identity matrix, $\|\cdot\|$ denotes the Euclidean vector norm or corresponding matrix norm, and $A\succ0$ means that the symmetric matrix $A$ is positive definite. For a signal $u(t)$, a starting time $t_0\ge0$, a terminal time $t_f>t_0$, and a constant $\beta\ge0$, define its exponentially weighted $L_2$ norm by
\begin{equation}\label{eq:weighted_norm}
	\begin{split}
&\|u\|_{2,\beta;[t_0, t_f]} =\left(\int_{t_0}^{t_f}
 e^{2\beta(t-t_0)}\|u(t)\|^2\,dt\right)^{1/2}
 \\
&\text{and}\quad \|u\|_{2,\beta;t_0} =\left(\int_{t_0}^{\infty}
e^{2\beta(t-t_0)}\|u(t)\|^2\,dt\right)^{1/2}.
\end{split}
\end{equation}
For a causal dynamic system $\mathcal S$ with input $u$ and output $y=\mathcal Su$, its \emph{weighted induced $L_2$ gain} is denoted by 
\begin{equation}\label{eq:induced_gain}
\|\mathcal S\|_{2,\beta}=\sup_{t_0\ge0}\ \sup_{u\ne0}
\frac{\|y\|_{2,\beta;t_0}}{\|u\|_{2,\beta;t_0}},
\end{equation}
where the system starts from zero initial conditions at $t_0$, including zero past input for a system with delay. The supremum is taken over inputs with finite, nonzero weighted $L_2$ norm. Thus $\|\mathcal S\|_{2,\beta}$ is the largest input-to-output gain, measured uniformly over starting times. At $\beta=0$, this reduces to the standard unweighted induced $L_2$ gain.

\section{Problem and Objective Statement}\label{sec:problem}
Consider a nonlinear system
\begin{equation}\label{eq:plant}
\dot x(t)=f(x(t),u(t),t)+\Phi(t)\theta(t),
\end{equation}
where $x(t)\in\R^n$ is the system state, $u(t)\in\R^m$ is the known control input, $f(x,u,t)\in\R^n$ represents the known part of the system dynamics, $\Phi(t)\in\R^{n\times p}$ is the known regressor, and $\theta(t)\in\R^p$ is the unknown time-varying parameter to be estimated. The regressor may depend on measured states, known inputs, and time. The unknown parameters are assumed twice continuously differentiable and there exists a positive constant $\bar d$ such that $\|\ddot\theta(t)\|\le\bar d$, but their values and derivatives including $\bar d$ are not used by the learner's design and implementation.

\begin{assumption}\label{ass:regressor}
The known regressor $\Phi(t)$ and its first time derivative are uniformly bounded. In particular, $\|\Phi(t)\|\le\bar\Phi$ for some constant $\bar\Phi>0$. Furthermore, there exist $T,\mu>0$ such that
\begin{equation}\label{eq:pe}
\int_t^{t+T}\Phi^\top(s)\Phi(s)\,ds\succeq\mu I_p,\qquad t\ge0.
\end{equation}
\end{assumption}

The boundedness condition limits the variation of the available signals, while the PE condition~\eqref{eq:pe} ensures that the measured data provide sufficiently rich information about every unknown parameter over each time interval of length $T$. Neither condition restricts the unknown parameters to be constant or slowly varying.

The objective of this paper is to design an adaptive learner that estimates both $\theta(t)$ and its instantaneous rate $\dot\theta(t)$, guarantees stable learning under Assumption~\ref{ass:regressor}, and makes the estimation error depend on the second derivative of $\theta(t)$ rather than its first derivative. The method should use measured $x,u$, known $f,\Phi$, and stored parameter estimates only.

\section{Main Results}\label{sec:results}
We propose the following delay-enabled adaptive learning law:
\begin{subequations}\label{eq:learner}
\begin{align}
\dot{\hat x}&=f(x,u,t)+\Phi(t)\hat\theta-K(\hat x-x),\label{eq:pred}\\
\dot{\hat\theta}&=-\Gamma\Phi^\top(t)(\hat x-x)+\hat v,\label{eq:adapt}\\
\tau\dot{\hat v}&=-\hat v+\frac{\hat\theta(t)-\hat\theta(t-h)}h, \label{eq:rate}
\end{align}
\end{subequations}
where $\hat x\in\R^n$ is the state prediction, $\hat\theta,\hat v\in\R^p$ are the estimates of the unknown parameters and their rates, $K\in\R^{n\times n}$ and $\Gamma\in\R^{p\times p}$ are constant symmetric positive-definite design matrices, and $\tau,h>0$ are scalar design constants. $h$ is the introduced delay; $\hat\theta(s)$ on $[-h,0]$ is a prescribed continuous initial estimate history.
Note that the above adaptive learning  structure retains the state predictor~\eqref{eq:pred} and the gradient correction \eqref{eq:adapt} from conventional adaptive estimation architectures (see, e.g., \cite{ioannou}). The only addition to the parameter update is the rate estimate $\hat v$, generated by the first-order filter~\eqref{eq:rate} using the delayed difference of $\hat\theta$. Thus, learning the parameter variation requires only one additional dynamic state and a stored estimate history of length $h$.

For analysis, define the estimation errors
\begin{equation}\label{eq:errorsdef}
e_x=\hat x-x,\qquad e_\theta=\hat\theta-\theta,
\qquad e_v=\hat v-\dot\theta.
\end{equation}
Differentiating \eqref{eq:errorsdef} and using \eqref{eq:plant}--\eqref{eq:learner} gives the following error dynamics:
\begin{subequations}\label{eq:error}
\begin{align}
\dot e_x&=-Ke_x+\Phi(t)e_\theta,\label{eq:ex}\\
\dot e_\theta&=-\Gamma\Phi^\top(t)e_x+e_v,\label{eq:etheta}\\
\tau\dot e_v&=-e_v+\frac{e_\theta(t)-e_\theta(t-h)}h+w(t),\label{eq:ev}
\end{align}
\end{subequations}
where the perturbation term $w(t)$ is
\begin{equation}\label{eq:dist}
w(t)=\frac{\theta(t)-\theta(t-h)}h
-\dot\theta(t)-\tau\ddot\theta(t).
\end{equation}
Note that Equations~\eqref{eq:ex}--\eqref{eq:etheta} form the conventional predictor--adaptation subsystem in a linear time-varying (LTV) form driven by $e_v$; Equation~\eqref{eq:ev} is the delay-based rate-learning subsystem driven by $e_\theta$ and $w$. The two subsystems are interconnected through $e_\theta$ and $e_v$. Based on this observation, the following theorem establishes exponential stability of such an interconnected system using the small-gain theorem \citep{vanderschaft2017}. To this end, denote by $\mathcal S_a$ the subsystem~\eqref{eq:ex}--\eqref{eq:etheta} with input $e_v$ and output $e_\theta$; denote by $\mathcal S_d$ the subsystem~\eqref{eq:ev} with $w=0$, input $e_\theta$, and output $e_v$. The gains of both subsystems use the definition in~\eqref{eq:induced_gain}.


\begin{theorem}\label{thm:main}
Under Assumption~\ref{ass:regressor}, for any given $K,\Gamma\succ0$ and $h>0$, there exist $\tau>0$ and $\beta>0$, with $\beta\tau<1$, such that
\begin{equation}\label{eq:sg}
\|\mathcal S_a\|_{2,\beta}\,\|\mathcal S_d\|_{2,\beta}<1.
\end{equation}
For any parameters satisfying~\eqref{eq:sg}, the error system~\eqref{eq:error} is exponentially stable at the origin when $\ddot\theta(t)=0$ on $[-h,\infty)$. If $\theta(t)$ is twice continuously differentiable on $[-h,\infty)$ and $\|\ddot\theta(t)\|\le\bar d$, the errors approach a neighborhood of zero exponentially and satisfy
\begin{equation}\label{eq:ultimate}
\limsup_{t\to\infty}\left\|\begin{bmatrix}e_x(t)\\e_\theta(t)\\e_v(t)\end{bmatrix}\right\|
\le\frac{M_c}{\beta}\left(1+\frac h{2\tau}\right)\bar d,
\end{equation}
for some constant $M_c>0$.
\end{theorem}
\begin{proof}
	The proof proceeds in three steps. 
	
\emph{Step 1:} 
We first establish that condition~\eqref{eq:sg} can be satisfied. For the conventional predictor--adaptation subsystem $\mathcal S_a$ \eqref{eq:ex}--\eqref{eq:etheta}, set
\begin{equation}\label{eq:zaAB}
 z_a=\begin{bmatrix}e_x\\e_\theta\end{bmatrix},\quad
 A(t)=\begin{bmatrix}-K&\Phi(t)\\-\Gamma\Phi^\top(t)&0\end{bmatrix},\quad
 B=\begin{bmatrix}0\\I_p\end{bmatrix}.
\end{equation}
Then we have $\dot z_a=A(t)z_a+Be_v$, which is in the standard LTV form as considered in \cite{yuanW2011}. More specifically, under Assumption~\ref{ass:regressor}, the standard LTV stability result of \cite{yuanW2011} proves that there exist some constants  $M\ge1$ and $\lambda>0$ such that  
\begin{equation}\label{eq:peconv}
\|\Psi(t,s)\|\le Me^{-\lambda(t-s)},\qquad t\ge s\ge0, 
\end{equation}
for the subsystem $\mathcal S_a$, where $\Psi(t,s)$ is the associated state-transition matrix. 
The variation-of-constants formula~\citep{fridmanBook} further yields 
\begin{equation}\label{eq:voc}
 z_a(t)=\Psi(t,0)z_a(0)+\int_0^t\Psi(t,s)Be_v(s)\,ds.
\end{equation}
Using~\eqref{eq:peconv} and Young's convolution inequality, we obtain $\|\mathcal S_a\|_{2,\beta}\le M/(\lambda-\beta)$ for $0<\beta<\lambda$. For the delay subsystem $\mathcal S_d$ \eqref{eq:ev}, the transfer function from $e_\theta$ to $e_v$ with $w=0$ and zero initial conditions is $(1-e^{-sh})/[h(\tau s+1)]I_p$. Its weighted induced $L_2$ gain satisfies the following bound:   
\[
 \|\mathcal S_d\|_{2,\beta}\le \frac{e^{\beta h}}\tau
 \quad\text{when } 0<\beta\tau\le\tfrac12.
\]
To see this bound, use $|1-e^{(\beta-j\omega)h}|\le h e^{\beta h}\sqrt{\beta^2+\omega^2}$ and $\tau\sqrt{\beta^2+\omega^2}\le \sqrt{(1-\beta\tau)^2+\tau^2\omega^2}$ when $\beta\tau\le1/2$. Set $\beta=1/(4\tau)$. For fixed $K,\Gamma,h$, choose $\tau$ to ensure $\beta<\lambda$ and
\[
\|\mathcal S_a\|_{2,\beta}\|\mathcal S_d\|_{2,\beta}
\le \frac{M e^{h/(4\tau)}}{\tau(\lambda-1/(4\tau))}<1,
\]
since the right-hand side tends to zero as $\tau\to\infty$. This proves the stated existence of suitable parameters to satisfy condition \eqref{eq:sg}.

\emph{Step 2:} We next prove exponential stability for any parameters satisfying~\eqref{eq:sg} under $\ddot\theta=0$. 
By~\eqref{eq:dist}, it is straightforward that $\ddot\theta(t)=0$ on $[-h,\infty)$ implies $w(t)=0$ for $t\ge0$. Note that the two gains in~\eqref{eq:sg} bound the effect of $e_v$ on $e_\theta$ and the effect of $e_\theta$ on $e_v$, respectively, under zero initial conditions. To account for the actual initial errors, let $t_0\ge0$ be any starting time and define
\[
 D_{t_0}=\|z(t_0)\|+
 \sup_{t_0-h\le s\le t_0}\|e_\theta(s)\|,
 \qquad z=\begin{bmatrix}e_x\\e_\theta\\e_v\end{bmatrix}.
\]
The initial response of \eqref{eq:ex}--\eqref{eq:etheta} has a finite weighted $L_2$ norm by \eqref{eq:peconv} provided that $\beta<\lambda$. More precisely, the contribution of
$z_a(t_0)$ to $e_\theta$ is bounded, in the weighted norm over
$[t_0,t_f]$, by
$M\|z_a(t_0)\|/\sqrt{2(\lambda-\beta)}$.
The dynamics of \eqref{eq:ev} has a homogeneous response
$e^{-(t-t_0)/\tau}e_v(t_0)$, whose weighted $L_2$ norm is bounded by
$\|e_v(t_0)\|/\sqrt{2(1/\tau-\beta)}$ because $\beta\tau<1$.
Its delayed input over $[t_0,t_0+h]$ also depends on the prescribed
history $e_\theta$ on $[t_0-h,t_0]$. The weighted $L_2$ norm of the
corresponding input $-e_\theta(t-h)/h$ over $[t_0,t_0+h]$ is at most
$e^{\beta h}\sup_{t_0-h\le s\le t_0}\|e_\theta(s)\|/\sqrt{h}$.
Since the weighted $L_2$ gain of the first-order filter
$1/(\tau s+1)$ is $1/(1-\beta\tau)$, the resulting history
contribution is finite and proportional to $D_{t_0}$.
Consequently, for every $t_f>t_0$,
\begin{align*}
	\|e_\theta\|_{2,\beta;[t_0,t_f]}
	&\le C_aD_{t_0}+\|\mathcal S_a\|_{2,\beta}\,
	\|e_v\|_{2,\beta;[t_0,t_f]},\\
	\|e_v\|_{2,\beta;[t_0,t_f]}
	&\le C_dD_{t_0}+\|\mathcal S_d\|_{2,\beta}\,
	\|e_\theta\|_{2,\beta;[t_0,t_f]},
\end{align*}
where $C_a =\frac{M}{\sqrt{2(\lambda-\beta)}}$ and $C_d=\frac{1}{\sqrt{2(1/\tau-\beta)}}
		+\frac{e^{\beta h}}{\sqrt{h}(1-\beta\tau)}$ are constants independent of $t_0$ and $t_f$.
Substituting the second inequality into the first yields $\bigl(1-\|\mathcal S_a\|_{2,\beta}\|\mathcal S_d\|_{2,\beta}\bigr)
	\|e_\theta\|_{2,\beta;[t_0,t_f]}
	\le
	\bigl(C_a+\|\mathcal S_a\|_{2,\beta}C_d\bigr)D_{t_0}$. 
Condition~\eqref{eq:sg} guarantees that the coefficient on the left is
positive. Hence, we have
\begin{align*}
	\|e_\theta\|_{2,\beta;[t_0,t_f]}
	&\le
	\frac{C_a+\|\mathcal S_a\|_{2,\beta}C_d}
	{1-\|\mathcal S_a\|_{2,\beta}\|\mathcal S_d\|_{2,\beta}}D_{t_0},\\
	\|e_v\|_{2,\beta;[t_0,t_f]}
	&\le
	\frac{C_d+\|\mathcal S_d\|_{2,\beta}C_a}
	{1-\|\mathcal S_a\|_{2,\beta}\|\mathcal S_d\|_{2,\beta}}D_{t_0}.
\end{align*}
Furthermore, applying~\eqref{eq:peconv} to~\eqref{eq:voc} with initial time
$t_0$, and using Young's convolution inequality gives
\begin{align*}
	\|z_a\|_{2,\beta;[t_0,t_f]}
	&\le \frac{M}{\sqrt{2(\lambda-\beta)}}\|z_a(t_0)\| +\frac{M}{\lambda-\beta}
	\|e_v\|_{2,\beta;[t_0,t_f]}\\
	&\le \left[C_a+\frac{M}{\lambda-\beta}
	\frac{C_d+\|\mathcal S_d\|_{2,\beta}C_a}
	{1-\|\mathcal S_a\|_{2,\beta}\|\mathcal S_d\|_{2,\beta}}\right]D_{t_0}.
\end{align*}
Since $z=\mathrm{col}(z_a,e_v)$, the triangle inequality now yields
$\|z\|_{2,\beta;[t_0,t_f]}\le CD_{t_0}$,
where $C =C_a+\left(1+\frac{M}{\lambda-\beta}\right)
	\frac{C_d+\|\mathcal S_d\|_{2,\beta}C_a}
	{1-\|\mathcal S_a\|_{2,\beta}\|\mathcal S_d\|_{2,\beta}}$. 
Crucially, $C$ does not depend on $t_f$. Taking $t_f\to\infty$
and using monotone convergence for the nonnegative integrand,
we obtain
\begin{equation}
	\|z\|_{2,\beta;t_0} = \left(\int_{t_0}^{\infty}e^{2\beta(t-t_0)}\|z(t)\|^2\,dt\right)^{1/2} \le CD_{t_0}.
	\label{eq:weighted_full_bound}
\end{equation}

We now show why this integral bound also gives exponential decay at each time.
For $t\ge t_0+h$, averaging the identity
$z(t)=z(s)+\int_s^t\dot z(r)\,dr$ over $s\in[t-h,t]$
and applying Cauchy--Schwarz yields
\begin{equation}\label{eq:pointwise_bridge}
	\begin{aligned}
		\|z(t)\|\le{}&\frac{1}{\sqrt h}
		\left(\int_{t-h}^t\|z(s)\|^2\,ds\right)^{1/2}\\
		&+\sqrt h\left(\int_{t-h}^t\|\dot z(s)\|^2\,ds\right)^{1/2}.
	\end{aligned}
\end{equation}

Since $\Phi(t)$ is uniformly
bounded, the error dynamics \eqref{eq:error} with $w=0$ imply that there
exists a constant $L>0$ such that
$
\|\dot z(s)\|
\le L\bigl(\|z(s)\|+\|e_\theta(s-h)\|\bigr)$.  In particular, for $s\ge t_0+h$, this gives
$\|\dot z(s)\|
\le L\bigl(\|z(s)\|+\|z(s-h)\|\bigr)$.
For $t\ge t_0+2h$, we first use \eqref{eq:weighted_full_bound} to obtain
\begin{align*}
	\int_{t-h}^{t}\|z(s)\|^2\,ds
	&\le e^{-2\beta(t-t_0-h)}
	\int_{t-h}^{t}
	e^{2\beta(s-t_0)}\|z(s)\|^2\,ds\\
	&\le C^2D_{t_0}^2 e^{-2\beta(t-t_0-h)}.
\end{align*}
Consequently,
$
\left(\int_{t-h}^{t}\|z(s)\|^2\,ds\right)^{1/2}
\le CD_{t_0}e^{-\beta(t-t_0-h)}$. 

Next, by the bound on $\dot z$ and Minkowski's inequality,
\begin{align*}
	\left(\int_{t-h}^{t}\|\dot z(s)\|^2\,ds\right)^{1/2}
	&\le
	L\left(\int_{t-h}^{t}\|z(s)\|^2\,ds\right)^{1/2}\\
	&\quad+
	L\left(\int_{t-h}^{t}\|z(s-h)\|^2\,ds\right)^{1/2}.
\end{align*}
Using the change of variable $r=s-h$, the second integral
becomes
$
\int_{t-h}^{t}\|z(s-h)\|^2\,ds
=
\int_{t-2h}^{t-h}\|z(r)\|^2\,dr$. 
Applying \eqref{eq:weighted_full_bound} to both intervals yields
\begin{align*}
	\left(\int_{t-h}^{t}\|\dot z(s)\|^2\,ds\right)^{1/2}
	&\le LCD_{t_0}
	\left[
	e^{-\beta(t-t_0-h)}
	+e^{-\beta(t-t_0-2h)}
	\right]\\
	&\le 2LCD_{t_0}e^{-\beta(t-t_0-2h)}.
\end{align*}

Substituting the preceding results  into~\eqref{eq:pointwise_bridge}, we obtain
\begin{align*}
	\|z(t)\|
	&\le \frac{CD_{t_0}}{\sqrt h}
	e^{-\beta(t-t_0-h)}
	+2\sqrt h\,LCD_{t_0}
	e^{-\beta(t-t_0-2h)}\\
	&\le C\left(
	\frac{e^{\beta h}}{\sqrt h}
	+2\sqrt h\,L e^{2\beta h}
	\right)
	D_{t_0}e^{-\beta(t-t_0)}.
\end{align*}
This establishes pointwise exponential decay for
$t\ge t_0+2h$.

For the remaining interval $t_0\le t\le t_0+2h$,
let $m(t)=\max\bigl\{\sup_{t_0-h\le s\le t_0}\|e_\theta(s)\|,\,\sup_{t_0\le s\le t}\|z(s)\|\bigr\}$.
The bound on $\dot z$ and the initial history imply
$m(t)\le D_{t_0}+2L\int_{t_0}^t m(s)\,ds$.
Gronwall's inequality therefore gives
$\|z(t)\|\le D_{t_0}e^{2L(t-t_0)}
\le e^{(4L+2\beta)h}
D_{t_0}e^{-\beta(t-t_0)}$. 
Combining the two time intervals, we conclude that
\begin{equation}\label{eq:expcomplete}
	\|z(t)\|\le M_c e^{-\beta(t-t_0)}D_{t_0},\qquad t\ge t_0,
\end{equation}
where $M_c=\max\left\{
e^{(4L+2\beta)h},
C\left(
\frac{e^{\beta h}}{\sqrt h}
+2\sqrt h\,L e^{2\beta h}
\right)
\right\}$.
Thus the complete error system \eqref{eq:error} is uniformly exponentially
stable when $\ddot\theta=0$.

\emph{Step 3:} Lastly, we prove the error bound for bounded $\ddot\theta$. 
The only external perturbation of \eqref{eq:error} is $w$. We first bound it directly from~\eqref{eq:dist}. The fundamental theorem of calculus gives
$
\theta(t)-\theta(t-h)=\int_{t-h}^t\dot\theta(s)\,ds$ and $\dot\theta(s)=\dot\theta(t)-\int_s^t\ddot\theta(r)\,dr$. 
Substituting the second identity into the first and exchanging the order of integration yields
$
\theta(t)-\theta(t-h)
 =h\dot\theta(t)-\int_{t-h}^t(r-t+h)\ddot\theta(r)\,dr$. 
It follows from~\eqref{eq:dist} that
$
 w(t)=-\tau\ddot\theta(t)
 -\frac1h\int_{t-h}^t(r-t+h)\ddot\theta(r)\,dr$. 
For $r\in[t-h,t]$, the weight $r-t+h$ is nonnegative. Therefore, using $\|\ddot\theta(r)\|\le\bar d$ and the triangle inequality,
\begin{align*}
\|w(t)\|&\le\tau\bar d
 +\frac{\bar d}{h}\int_{t-h}^t(r-t+h)\,dr=\left(\tau+\frac h2\right)\bar d.
\end{align*}

The exponential estimate~\eqref{eq:expcomplete} is uniform over starting times and includes the initial history. Thus, the variation-of-constants formula for linear time-delay systems~\citep{fridmanBook} bounds the forced response by convolution with an exponentially decaying kernel. Since $w$ enters only the $e_v$ equation as $w/\tau$, the input matrix has norm $1/\tau$. Enlarging $M_c$ if necessary, without changing~\eqref{eq:expcomplete}, gives the following bound:
$
\|z(t)\|\le M_c e^{-\beta t}D_{0}
 +\frac{M_c}{\tau}\int_0^t e^{-\beta(t-s)}\|w(s)\|\,ds
$. 
Using the preceding bound on $w$ and integrating the exponential kernel, we obtain
\[
\|z(t)\|\le M_c e^{-\beta t}D_{0}
 +\frac{M_c}{\beta}\left(1+\frac h{2\tau}\right)
 (1-e^{-\beta t})\bar d.
\]
The first term decays exponentially, while the second is bounded by the right-hand side of~\eqref{eq:ultimate}. This completes the proof. 
\end{proof}

\begin{remark}\label{rem:exact}
The exponential-stability conclusion of Theorem~\ref{thm:main} implies that the state-prediction error, parameter estimation error, and parameter-rate estimation error all converge to zero exponentially when $\ddot\theta(t)=0$. An important consequence is that the proposed scheme can accurately estimate both the instantaneous values and the constant rates of linearly time-varying parameters of the form $\theta(t)=\theta_0+vt$, where $\theta_0$ and $v$ are unknown constants. 
\end{remark}

\begin{remark}\label{rem:bounded}
The bounded-error conclusion of Theorem~\ref{thm:main} addresses parameters whose rates also change with time. When $\ddot\theta$ is bounded, the three estimation errors approach a bounded neighborhood of zero. The bound in~\eqref{eq:ultimate} is proportional to the upper bound $\bar d$ on $\|\ddot\theta\|$: smaller changes in the parameter rate lead to a smaller guaranteed error bound. Thus a parameter can change quickly while its estimation error remains small, provided its rate changes slowly. 
\end{remark}

\begin{remark}\label{rem:verification}
The small-gain condition~\eqref{eq:sg} is sufficient, but not necessary, for the stated stability and error bounds. In implementation, one may choose $K$, $\Gamma$, $h$, and $\tau$, then evaluate the two weighted induced $L_2$ gains for the respective interconnected subsystems. If condition \eqref{eq:sg} is not met, these design parameters can be adjusted and the condition checked again. More detailed illustrations will be provided in   Section~\ref{sec:sim} using specific examples.
\end{remark}

\section{Numerical Studies}\label{sec:sim}
To illustrate the proposed theoretical results, we consider the following nonlinear system:
\begin{equation*}
\dot x=\begin{bmatrix}x_2\\-1.2x_1-0.8x_2-0.15x_1^3+u\end{bmatrix}
+\Phi(t)\theta(t),
\end{equation*}
where $u(t)=0.6\sin(0.6t)+0.4\cos(1.1t)$, with 
\begin{equation}\label{eq:phi}
\Phi(t)=\frac{g(t)}{\sqrt{2}}
\begin{bmatrix}1&1\\-1&1\end{bmatrix},\quad
 g(t)=\begin{cases}
\frac{1-\cos(2\pi t)}{2},&0\le t<1,\\
0,&1\le t<2.
\end{cases}
\end{equation}
Here $g(t+2)=g(t)$ for all $t\ge0$, so $g$ is extended periodically beyond $[0,2)$. Equation \eqref{eq:phi} indicates that excitation information is unavailable during the second half of each two-second period, but the regressor and its derivative are bounded and
\begin{equation*}
\int_t^{t+2}\Phi^\top(\ell)\Phi(\ell)\,d\ell
=\frac38 I_2,\qquad t\ge0.
\end{equation*}
Thus Assumption~\ref{ass:regressor} is satisfied. 
For the estimator's parameters, we set $K=10I_2$, $\Gamma=150I_2$, $h=0.55$, and $\tau=1.10$, and then check~\eqref{eq:sg} with $\beta=0.01$. The two subsystem gains are computed as follows.

For the {predictor--adaptation subsystem \eqref{eq:ex}--\eqref{eq:etheta},} with the above given system setup, an upper bound on the weighted induced $L_2$ gain of $\mathcal S_a$ from $e_v$ to $e_\theta$ can be computed using the following periodic Lyapunov matrix inequality, yielding $\|\mathcal S_a\|_{2,0.01}\le\gamma = 1.23$ numerically: 
\begin{equation*}
\begin{bmatrix}
\dot P+(A+\beta I_4)^\top P+P(A+\beta I_4)+C^\top C&PB\\
B^\top P&-\gamma^2 I_2
\end{bmatrix}\prec0, 
\end{equation*}
where  $A$ and $B$ are defined in \eqref{eq:zaAB}, $C=\begin{bmatrix}0&I_2\end{bmatrix}$ so that $e_\theta=Cz_a$, and $P(t)\succ0$ is a periodic symmetric Lyapunov matrix. 
The above inequality ensures that the quadratic storage $V=z_a^\top P(t)z_a$ satisfies $\dot V+2\beta V+\|e_\theta\|^2-\gamma^2\|e_v\|^2<0$; integration with zero initial conditions gives $\|e_\theta\|_{2,\beta;t_0}\le\gamma\|e_v\|_{2,\beta;t_0}$. Thus the resulting $\gamma=1.23$ is a gain upper bound, rather than necessarily the exact induced gain. 

For the {delay subsystem \eqref{eq:ev}}, its transfer function is $G_d(s)=(1-e^{-sh})/[h(\tau s+1)]I_2$. The weighted induced $L_2$ gain of the delay subsystem is given by 
\begin{equation*}
\|\mathcal S_d\|_{2,\beta}
=\sup_{\omega\in\R}
\frac{|1-e^{\beta h}e^{-j\omega h}|}
{h\sqrt{(1-\beta\tau)^2+\tau^2\omega^2}}, 
\end{equation*}
we obtain $\|\mathcal S_d\|_{2,0.01}<0.8$ with the given $h=0.55$, $\tau=1.10$, and $\beta=0.01$. Consequently, the loop gain is bounded by $1.23\times0.8=0.984<1$, satisfying the small-gain condition \eqref{eq:sg}. 

The proposed delay-enabled adaptive learner is compared with two representative estimators: the conventional adaptive estimator (Conventional) without delay, using the same $K$ and $\Gamma$, and the time-varying learning-gain estimator of \citet{na2021} (Na-based, Eq.~(20) therein). For the Na-based estimator, the parameter update law and time-varying gain are
\begin{equation*}
\dot{\hat\theta}_{\mathrm{Na}}=-\gamma_{\mathrm{Na}}Q(t)(P_1+\kappa P_2),\quad
Q(t)=\bigl(e^{-\delta t}Q_0+G(t)\bigr)^{-1},
\end{equation*}
where $G$, $P_1$, and $P_2$ are defined by the filtered regressors in Eqs.~(2) and (9)--(11) of \citet{na2021}, and $Q_0=Q^{-1}(0)\succ0$. The filter time constant is $0.025$~s, $\gamma_{\mathrm{Na}}=3$, $\kappa=0.3$, and $Q_0=I_2$; the forgetting rate is $\delta=20$ for the first experiment and $\delta=15$ for the second, as will be detailed in the sequel. The plant starts at $x(0)=[0.5,-1]^\top$; zero initial conditions are set for all the estimators' states, including zero initial rate estimate and zero prehistory for the proposed estimator. We consider the following two scenarios of time-varying parameters for comparison: linearly time-varying parameters and smooth time-varying parameters with continuously changing parameter rates.

\subsection{Linearly time-varying parameters}
\begin{figure}[t]\centering\includegraphics[width=\columnwidth]{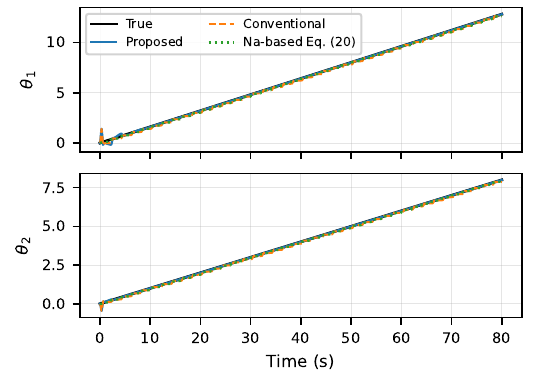}\caption{Learning of linearly time-varying parameters.}\label{fig:linearest}\end{figure}

\begin{figure}[t]
\centering\includegraphics[width=\columnwidth]{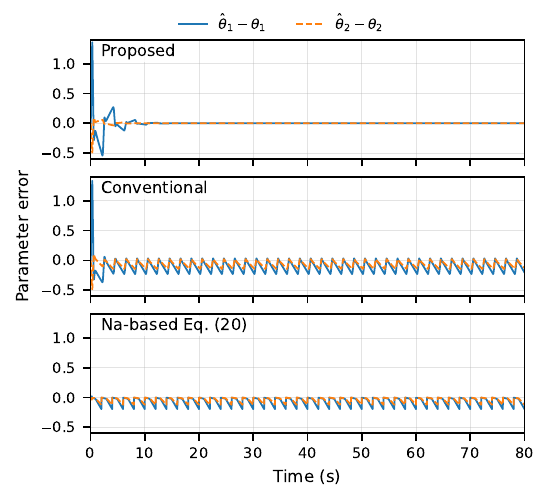}
\caption{Parameter estimation errors for linearly time-varying parameters. Each panel shows both parameter errors for one estimator, using a common vertical scale.}
\label{fig:linearerr}
\end{figure}

\begin{figure}[t]\centering\includegraphics[width=\columnwidth]{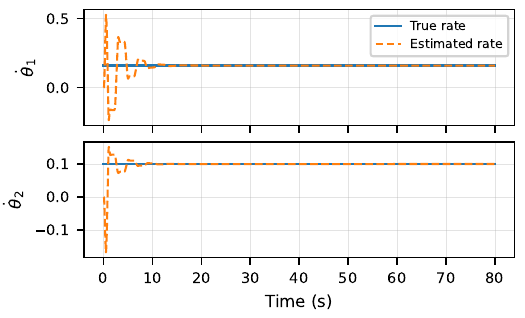}\caption{Learning of linearly time-varying parameter rates under the proposed method.}\label{fig:linearrate}\end{figure}

Consider
\begin{equation*}
\theta(t)=\begin{bmatrix}0.16t\\0.10t\end{bmatrix}.
\end{equation*}
The two unknown parameters change at constant rates. Since their second derivatives are zero, Theorem~\ref{thm:main} guarantees exponentially vanishing parameter and rate-estimation errors under the verified small-gain condition with the proposed method. Figures~\ref{fig:linearest}--\ref{fig:linearrate} compare the learning performance of the three estimators and illustrate the learned parameter rates.
In particular, Figure \ref{fig:linearerr} directly compares the errors in estimating both unknown parameters. For the
proposed learner, both errors decay rapidly to essentially zero after the initial transient and remain negligible even during
subsequent intervals without parameter excitation. The conventional and Na-based estimators instead exhibit persistent
periodic errors. Table \ref{tab:linear_rmse} confirms this behavior: over
$40$--$80$ s, the proposed learner achieves the lowest parameter-vector root-mean-square (RMS) error of $2.08\times10^{-8}$, compared with $0.15512$ (conventional) and $0.10455$ (Na-based). The proposed learner also attains the smallest RMS error over the full $0$--$80$ s interval among the three methods.
Figure \ref{fig:linearrate} further shows that under the proposed learner, both estimated rates approach their true constant values after a short transient and remain close to those values over the remainder of the simulation. These results support the exact-learning conclusion of Theorem \ref{thm:main} of the proposed delay-enabled adaptive learner for
linearly time-varying parameters.
\begin{table}[t]\centering\caption{Parameter-vector RMS errors for linearly time-varying parameters.}\label{tab:linear_rmse}\small\setlength{\tabcolsep}{4pt}
\begin{tabular}{lrr}\toprule
Method & $0$--$80$ s & $40$--$80$ s\\\midrule
Proposed & \textbf{0.09515} & $\boldsymbol{2.08\times 10^{-8}}$\\
Conventional & 0.17332 & 0.15512\\
Na-based (varying gain) & 0.10454 & 0.10455\\
\bottomrule
\end{tabular}
\end{table}

\subsection{Continuously changing parameter rates}
\begin{figure}[t]\centering
\includegraphics[width=\columnwidth]{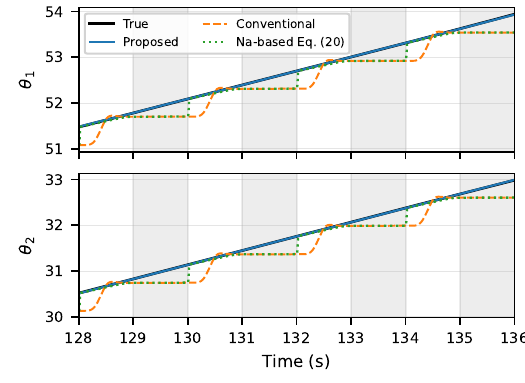}
\caption{Learning of parameters with continuously changing rates. Gray bands mark intervals where $\Phi(t)=0$.}\label{fig:intertrack}
\end{figure}

\begin{figure}[t]
\centering\includegraphics[width=\columnwidth]{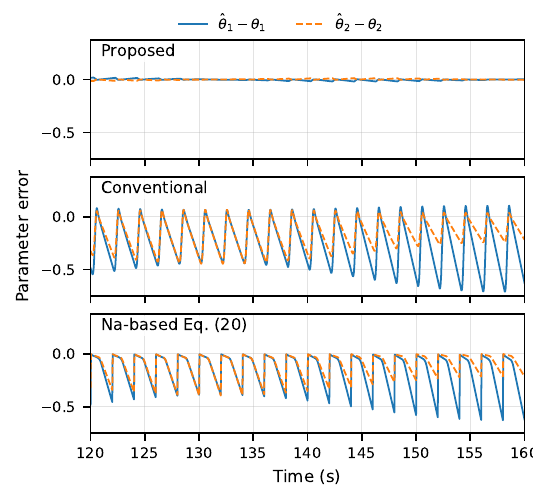}
\caption{Parameter estimation errors for parameters with continuously changing rates. Each panel shows both parameter errors for one estimator, using a common vertical scale.}
\label{fig:intererr}
\end{figure}

\begin{figure}[t]\centering
\includegraphics[width=\columnwidth]{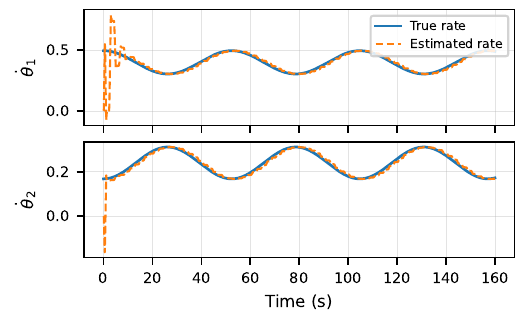}
\caption{Learning of continuously changing parameter rates under the proposed method.}\label{fig:interrate}
\end{figure}

In the second experiment, we consider parameters with continuously changing rates:
\begin{equation*}
\theta(t)=\begin{bmatrix}
0.4t+0.8\sin(0.12t)\\
0.24t-0.6\sin(0.12t)
\end{bmatrix}.
\end{equation*}
The rates now vary with time, so the parameter-estimation error need not converge exactly to zero. All three methods use the same regressor and the specified settings. Figures~\ref{fig:intertrack} and \ref{fig:intererr} display the tracking and error signals. Shaded bands indicate intervals without excitation information.
These plots show that the proposed method maintains both estimation errors near zero over
$120$--$160$ s, with small residual variations rather than convergence exactly to zero. In comparison, the conventional
method shows large repetitive error oscillations and the Na-based method retains substantial periodic errors.
Table \ref{tab:intermittent} confirms that the proposed method has the smallest RMS error among the three estimators over both the full $0$--$160$ s interval and the $120$--$160$ s interval, with late-time RMS errors of $0.00904$ (proposed), $0.38498$ (conventional), and $0.27753$ (Na-based).
Figure~\ref{fig:interrate} further shows that the estimated parameter rates closely follow their true time-varying values after the 
initial transient, including the sinusoidal changes in their values. Unlike the first experiment, the rate estimates here do not settle to constants. Small tracking discrepancies remain, as expected when the second derivatives of the parameters are nonzero.
Together, these results illustrate the distinction in Theorem~\ref{thm:main}: exact asymptotic learning for zero second derivatives and
bounded estimation errors for general smooth parameter variation. 

\begin{table}[t]\centering
\caption{Parameter-vector RMS errors for continuously changing rates.}
\label{tab:intermittent}\small\setlength{\tabcolsep}{4pt}
\begin{tabular}{lrr}\toprule
Method & $0$--$160$ s & $120$--$160$ s\\\midrule
Proposed & \textbf{0.07971} & \textbf{0.00904}\\
Conventional & 0.39382 & 0.38498\\
Na-based (varying gain) & 0.28077 & 0.27753\\\bottomrule
\end{tabular}
\end{table}

It is important to note from the above two experiments that, unlike conventional adaptive estimation, the proposed method explicitly retains a learned estimate of the parameter variation rate. The Na-based method uses additional filtered signals and a time-varying learning gain. The proposed method achieves substantially better learning performance in these two simulations while estimating both instantaneous parameter values and their variation rates. Its implementation requires only one delayed difference and a first-order rate-estimation state.

\section{Conclusion}\label{sec:conclusion}
This paper introduced a novel delay-enabled adaptive learning method for unknown time-varying parameters. By augmenting conventional adaptive estimation with a delayed parameter difference and a single first-order dynamic state, the proposed method enables simultaneous estimation of unknown parameter values and their variation rates. The resulting estimation errors depend on the second derivative of the unknown parameters rather than their first derivative. Under persistent excitation, exponential stability and explicit estimation-error bounds were established through a small-gain analysis. In particular, linearly time-varying parameters and their unknown constant rates can be learned accurately  with exponentially vanishing errors. Numerical comparisons with representative existing methods further demonstrated the effectiveness of the proposed learning mechanism.

The emphasis of this work has been on introducing this new learning capability and establishing its fundamental stability and convergence properties. Accordingly, the analysis relies on conventional PE results and the small-gain theorem, without resorting to more elaborate tools from time-delay systems theory, such as Lyapunov–Krasovskii functionals \citep{fridmanBook} or dynamic integral quadratic constraints \citep{yuanWIQCautomatica}. Although the resulting small-gain condition provides a compact stability criterion, designing learner parameters that achieve both stability and satisfactory learning performance may still require iterative tuning and repeated feasibility checks. Developing more systematic, efficient, and less conservative synthesis methods is therefore an important direction for future research. In particular, advanced time-delay analysis and optimization techniques may provide useful tools for this purpose, while retaining the simple delay-enabled learning architecture introduced in this paper.

\begingroup\footnotesize
\bibliographystyle{elsarticle-harv}
\bibliography{references}
\endgroup
\end{document}